\documentclass[letterpaper, 10 pt, conference]{cssconf}
\IEEEoverridecommandlockouts
\newtheorem{theorem}{Theorem}[section]

\newtheorem{lemma}[theorem]{Lemma}

\newtheorem{remark}[theorem]{Remark}
\newfont{\BB}{msbm10}
\newfont{\bb}{msbm8}
\def\cg{{\cal G}}
\def\ch{{\cal H}}

\def\tr{{\rm tr}}

\def\lim{{\rm lim}\ }
\def\span{{\rm span}\ }

\title{Randomized control of open quantum systems}

\author{ \parbox{3 in}{\centering Lorenza Viola \\
Department of Physics and Astronomy \\
Dartmouth College \\
6127 Wilder Laboratory  \\
Hanover, New Hampshire 03755, USA}\\
{\tt\small Lorenza.Viola@Dartmouth.edu}}

\begin{document}
 
\maketitle

\begin{abstract}  
The problem of open-loop dynamical control of generic open 
quantum systems is addressed.  In particular, I focus on the 
task of effectively switching off environmental couplings 
responsible for unwanted decoherence and dissipation effects. 
After revisiting the standard framework for dynamical decoupling
via deterministic controls, I describe a different approach 
whereby the controller intentionally acquires a random component.
An explicit error bound on worst-case performance of stochastic
decoupling is presented. 
\end{abstract}
\maketitle

\section{Introduction} 

The need for accurately controlling the dynamics of a
quantum-mechanical system is central to a variety of tasks 
ranging across contemporary physics, engineering, and information
sciences~\cite{blaquerie,brumer,NC}.  In particular, motivated by 
both continuous experimental advances in nanoscale devices and the
challenge to practically implement fault-tolerant quantum information
processing, control strategies for {\em open} quantum systems
undergoing realistic irreversible dynamics~\cite{BF} play an
increasingly prominent role.

Dynamical decoupling techniques offer a versatile control toolbox
for open quantum-system engineering~\cite{VL,VKL99,Viola02}. In 
its essence, a decoupling protocol consists in a sequence of 
{\em open-loop} transformations on the target system (control
pulses in the simplest setting), designed in such a way that 
the effect of unwanted dynamics is coherently averaged out in 
the resulting controlled evolution.  Applied to the removal of 
unwanted couplings between the target system and its surrounding
environment, this paves the way to a general strategy for 
decoherence control and error-suppressed quantum computation 
purely based on unitary control means. 

Both within formulations of the decoupling problem and more 
general coherent-control settings, the restriction to purely
{\em deterministic} control fields has provided a most natural
starting point.  In a way, this finds ample justification 
in the fact that non-deterministic effects (such as stochastic
noise and/or random control imperfections) typically deteriorate
system performance, motivating the effort for designing 
intrinsically robust decoupling schemes~\cite{VK} and for 
assessing open-loop fault-tolerance thresholds~\cite{kaveh}.  Yet, 
no fundamental reasons exist for not lifting such a restriction, by 
{\em purposefully allowing stochasticity} in the underlying control 
design. Beside being conceptually intriguing on its own, it is 
worth recalling that notable examples may be found of situations 
where noise and randomness might have a beneficial rather than 
detrimental effect.  Of special relevance are phenomena like the 
self-averaging of intermolecular interactions in gases and liquids 
via random microscopic motions~\cite{Haeberlen} and quantum stochastic 
resonance~\cite{SR}, or the idea of dissipation-assisted quantum
computation~\cite{beige}.

A first step toward exploring randomized quantum control was recently
taken by Viola and Knill~\cite{VK05}, confirming in principle the
possibility of enhanced system performance as compared to
deterministic control in relevant scenarios.  It is the purpose of
this paper to further elucidate the random decoupling framework, by
first presenting a general control-theoretic formulation and contrast
it to the standard deterministic one (Section II), and then discuss in
detail a quantitative error bound on stochastic control performance
(Section III). Final remarks conclude in Section IV.

\section{Formulation of the control problem}
\subsection{Quantum-control systems}

The standard {open-loop} control problem for an isolated, {\em
closed} quantum system $S$ defined on a state space $\ch_S$ of
dimension $d_S <\infty$ is described (in units where $\hbar=1$) by a
bilinear control system of the form~\cite{BS}
\begin{eqnarray}
{d U(t) \over dt}&\hspace*{-1mm}=\hspace*{-1mm}& 
-i \Big( H_0  + H_c (t) \Big) U(t)\:, 
\nonumber \\
H_c(t)&\hspace*{-1mm}=\hspace*{-1mm}&\sum_{\ell=1}^m H_\ell u_\ell (t)\:.  
\label{bil}
\end{eqnarray}
Here, $U(t)$  is the evolution  operator (or {\em propagator})  of the
system, whereas  $H_0\equiv H_S$, $H_\ell$ represent  the internal (or
{\em  drift})  Hamiltonian,  and  the  applied  control  Hamiltonians,
respectively.  Both $H_0$ and  the $H_\ell$ are Hermitian operators on
$\ch_S$  which, without  loss  of  generality, may  be  assumed to  be
traceless.   The time  dependence of  the overall  control Hamiltonian
$H_c(t)$  is modeled  through  the real  functions $u_\ell(t)$,  which
typically represent electromagnetic fields and are the control inputs
of the  problem.  A broad  separation between {\em  deterministic} and
{\em  stochastic} control systems  may be  drawn depending  on whether
each  control  input is  a  deterministic  function  of time  or  some
randomness is  allowed for at  least one input.   The state of  $S$ is
described  in general by  a Hermitian,  positive operator  $\rho_S$ on
$\ch_S$, normalized with respect to the  trace norm in such a way that
$\tr_S  (\rho_S)=1$.  In  what  follows,  I will  assume  that $S$  is
initially  in  a {\em  pure  state},  described  by a  one-dimensional
projector $\pi_S$ of the form $\pi_S=|\psi\rangle \langle \psi|$, with
$|\psi\rangle \in \ch_S$.

It is convenient to focus directly on the {\em control
propagator} $U_c(t)$ as the basic object for control design,
\begin{equation}
{U}_c(t)={\cal T} \hspace*{-.5mm}
\exp\left \{ -i \hspace{-0.5mm} \int_0^t  du 
{H}_c(u) \right\} \:, 
\end{equation}
where the symbol ${\cal T}$ denotes as usual time ordering.  By
effecting a canonical transformation to a time-dependent frame that
continuously follows the applied control,
\begin{equation}
\tilde{\rho}_S(t) = U^\dagger_c(t) \rho_S(t) U_c(t)\:,
\end{equation}
the explicit action of the control field is removed from the dynamics.
The control problem of Eq.~(\ref{bil}) takes the form
\begin{eqnarray}
{d \tilde{U}(t) \over dt}&\hspace*{-1mm}=\hspace*{-1mm}& 
-i \tilde{H}(t) \tilde{U}(t)\:, 
\nonumber \\
\tilde{H}(t)&\hspace*{-1mm}=\hspace*{-1mm}&
U_c^\dagger (t) H_0 U_c(t)\:,  
\label{log}
\end{eqnarray}
in terms of the propagator $\tilde{U}(t)$ for the transformed state,
\begin{equation}
\tilde{\rho}_S(t)= \tilde{U}(t) \tilde{\rho}_S(0) 
\tilde{U}^\dagger (t)\:, \hspace{3mm} 
\tilde{U}(t)={U}_c^\dagger (t) {U}(t)\:.
\end{equation}
I will refer to the formulations of Eqs.~(\ref{bil}), (\ref{log}) as
{\em physical} and {\em logical} frame formulations, respectively.
While from the mathematical point of view the logical frame
description has the disadvantage of being highly non-linear in the
control inputs, Eq.~(\ref{log}) makes it very convenient to directly
map properties of the desired effective evolution back into design
constraints for $U_c(t)$, and viceversa.  If the control strategy is
{\em cyclic}, that is $U_c(t+T_c)=U_c(t)$ for $T_c>0$, and $H_0$ is
time-independent as assumed so far, the periodicity of the control
field is transferred to the logical Hamiltonian $\tilde{H}(t)$, and an
exact representation of the controlled evolution in terms of {\em
average Hamiltonian theory} exists~\cite{VKL99,Haeberlen},
\begin{equation}
\tilde{U}(t)=e^{-i \overline{H}t}\:, \hspace{3mm}
\overline{H}=\sum_{\kappa=0}^\infty \overline{H}^{(\kappa)} \:,
\end{equation}
each term $\overline{H}^{(\kappa)}$ being computed from the Magnus
series for $\tilde{H}(t)$.  As it turns out, the logical formulation
is also particularly useful in situations where the control strategy
directly incorporates {\em symmetry} criteria.

For a realistic {\em open} quantum system, the influence of the
surrounding environment may modify the dynamics in two important ways.
$(i)$ $S$ may couple to a {\em classical} environment, effectively
resulting into a (possibly random) time-dependent modification of the
system parameters, in particular $H_S \mapsto H_S(t)$.  Deterministic
time-dependent quantum control systems have been recently investigated
in~\cite{Clark05}.  $(ii)$ $S$ may couple to a {\em quantum}
environment $E$, that is a second quantum subsystem defined on a state
space $\ch_E$ of dimension $d_E >>d_S$ and characterized by an
internal Hamiltonian $H_E$.  Let ${\bf I}_{S,E}$ denote the identity
operator on $\ch_{S,E}$, respectively.  The drift Hamiltonian
$H_0(t)\equiv H_{SE}(t)$ of a general open quantum system may then be
expressed as
\begin{equation}
H_0(t)=H_S(t) \otimes {\bf I}_E + {\bf I}_S \otimes H_E +
\sum_a J_a (t) \otimes B_a \:,
\label{totaldrift}
\end{equation}
where the $B_a$'s are linearly independent environment operators and,
without loss of generality, we may assume the coupling operators (or
{\em error generators}) to be traceless.  In typical situations, both
the exact time dependence of $H_S(t)$ and $J_a(t)$, as well as the
exact form of $H_E, B_a$ are unknown.  If $\rho_{SE}(t)$ denotes the
{\em joint} state of the composite $S, E$ system, the evolution of $S$
alone is now described by the {\em reduced} state obtained by a partial
trace over $E$,
\begin{equation}
\rho_S(t)= \tr_E ( \rho_{SE}(t)) \:.
\label{ptrace}
\end{equation}
In general, the evolution of an initially pure state $\pi_S$ of $S$
under the Hamiltonian (\ref{totaldrift}), followed by $(i)$ the
ensemble average over the resulting time histories and/or $(ii)$ the
partial trace (\ref{ptrace}), results in a {\em mixed} state of $S$,
$\tr(\rho_S^2(t))<1$.  This implies genuinely {\em non-unitary},
irreversible dynamics for $S$, which physically accounts for quantum
decoherence and dissipation effects~\cite{BF}.

For an open system, a control problem {\em formally} similar to
(\ref{bil}) may still be formulated for the combined propagator 
$U(t)$ of $S$ plus $E$, provided that the action of the controller
is explicitly restricted to the {\em system variables only} that is,
\begin{equation}
H_c(t) \equiv H_c(t)\otimes {\bf I}_E\:, \hspace{3mm} 
U_c(t) \equiv U_c(t)\otimes {\bf I}_E \:. 
\end{equation}
Two frame transformations may be relevant in the open system context.
The transformation to a logical frame, which explicitly removes the
applied control Hamiltonian, is effected as before,
\begin{equation}
\tilde{\rho}_{SE}(t) = U^\dagger_c(t) \rho_{SE}(t) U_c(t)\:,
\end{equation}
leading to a control problem formally similar to (\ref{log}), 
with 
\begin{eqnarray}
\tilde{H}_{SE} (t) &\hspace*{-1mm}=\hspace*{-1mm}&
\left[ U_c^\dagger (t) H_S(t) U_c(t)\right] \otimes {\bf I}_E +
{\bf I}_S \otimes H_E + \nonumber \\
 &\hspace*{-1mm}+\hspace*{-1mm}& 
\sum_a \left[ U_c^\dagger (t) J_a (t) U_c(t) \right] \otimes B_a   \:.  
\label{lham}
\end{eqnarray}
If a formulation which also removes the evolution due to $H_E$ is
needed, a simultaneous canonical transformation to a logical
interaction frame is effected on the environment variables,
\begin{equation} 
\tilde{\rho}'_{SE}(t) = U_E^\dagger(t) \tilde{\rho}_{SE}(t) U_E(t)
\:,\hspace{3mm} U_E(t)=e^{-i H_E t}\:.
\end{equation}
The corresponding propagator $\tilde{U}'(t)$ still satisfies an 
equation similar to (\ref{log}), where now
\begin{eqnarray}
\tilde{H}'_{SE} (t) &\hspace*{-1mm}=\hspace*{-1mm}&
\left[U_c^\dagger (t) H_S(t) U_c(t) \right]\otimes 
{\bf I}_E + \label{hp}  \\
 &\hspace*{-1mm}+\hspace*{-1mm}& 
\sum_a \left[U_c^\dagger (t) J_a (t) U_c(t)\right] 
\otimes \left[U_E^\dagger (t) B_a U_E (t) \right] \:.  
\nonumber 
\end{eqnarray}
The various propagators are related to each other as follows:
\begin{equation}
U(t)=U_c(t) \tilde{U}(t)= U_c(t)U_E(t)\tilde{U}'(t)\:.
\label{relation}
\end{equation}

\subsection{Control tasks and performance indicators}

A {\em dynamical control} problem may be regarded as a steering
problem for the evolution operator of the target system in the
appropriate frame.  For an open system, a task of critical importance
is decoherence control, which effectively requires the suppression of
the error generators $J_a(t)$.  In particular, a {\em decoupling
problem} consists in determining a control configuration $\{H_\ell,
u_\ell(t)\}$ such that for a given evolution time $T>0$ the joint
propagator factorizes e.g.,
\begin{equation}
\tilde{U}(T)= \tilde{X}_S (T) \otimes U_E(T)\:,
\label{dec0}
\end{equation}
in the logical frame, $\tilde{X}_S(T)$ being a unitary operator on
$S$.  Notice that Eq. (\ref{dec0}) implies decoupling in the physical
frame as well.  The simplest decoupling objective, on which I will
focus henceforth, corresponds to identity design on $S$ (the so-called
{\em no-op} gate in quantum computation terminology~\cite{NC}, or
complete decoupling or annihilation in decoupling
terminology~\cite{VKL99,Viola02}), whereby
\begin{equation}
\tilde{U}(T)=  {\bf I}_S \otimes U_E(T)\:.
\label{dec1}
\end{equation} 

If both $H_S$ and the $J_a$ are constant in time, and $U_c(t)$ is
periodic, then the logical Hamiltonian (\ref{lham}) is also periodic
and the above equation, once fulfilled at time $T=T_c$, remains valid
for arbitrary times $T_N=N T_c$, $N \in {\bf N}$.  Under these
conditions, the logical and physical frames overlap for every $N$, and
the controlled evolution reads as
\begin{equation}
\tilde{\rho}_S(T_N)=\rho_S(T_N)=\rho_S(0)=\pi_S=
|\psi\rangle\langle \psi|\:.
\end{equation}
Thus, arbitrary initial states of $S$ are {\em stroboscopically 
preserved} in both the logical and the physical frames. If either
$H_S$ or $J_a$ are time-varying, and/or the control strategy is 
{\em acyclic}, it is still meaningful to require that
\begin{equation}
\tilde{\rho}_S(T)=\rho_S(0)=\pi_S \:, \hspace{3mm} T>0, \;\forall \pi_S\:.
\label{state}
\end{equation}
For stochastic control, the above objective is further relaxed to 
{\em average state preservation in the logical frame} that is, 
\begin{equation}
{\bf E} \left\{ \tilde{\rho}_S(T)\right\}
=\rho_S(0)=\pi_S \:, \hspace{3mm} T>0\:,\;\forall \pi_S\:,
\label{avstate}
\end{equation}
with ${\bf E}\{\:\}$ denoting ensemble expectation.  Clearly, control
schemes involving random operations are intrinsically acyclic, the
control path practically never returning the system to the physical
frame.  If, however, the past control trajectory is recorded, this may
be exploited to bring the state of $S$ back to the physical frame at
any time if desired.

In order to quantify the accuracy of a given control procedure at
achieving the intended objective, suitable performance indicators are
needed.  Let $\pi^\perp_S={\bf I}_S-|\psi\rangle\langle \psi|$ denote
the orthogonal complement of $\pi_S$ in $\ch_S$. Then the above task
(\ref{avstate}) is achieved if and only if, on average, the logical
(reduced) state of the system has zero component along $\pi^\perp_S$
(irrespective of the state of the environment).  This naturally
suggests to consider, for each pure initial state $\pi_S$, the
following {\em a priori error probability},
\begin{equation}
\epsilon_T(\pi_S)={\bf E}\left\{ \tr_S \left( \pi^\perp_S 
\tilde{\rho}_S(T)\right)\right\}\:.
\label{psep}
\end{equation}
Note that $\epsilon_T(\pi_S)\geq 0$ for all $\pi_S$ follows from the
fact that both $\pi^\perp_S $ and $\tilde{\rho}_S(T)$ are Hermitian
semi-positive definite operators.  A {\em worst-case pure state error
probability} may then be defined by maximizing over pure states that
is,
\begin{equation}
\epsilon_T= \mbox{Max}_{\pi_S \in \ch_S} 
\left\{ \epsilon_T(\pi_S)\right\} \:.
\label{ep}
\end{equation}

\subsection{Control assumptions and group-theoretical design}

Control design is strongly influenced by the class of available
controls.  A particularly simple scenario is provided by so-called
{\em quantum bang-bang controls}~\cite{VL,VKL99}, whereby the control
inputs $u_\ell(t)$ are able to be turned on and off impulsively with
unbounded strength, so as to implement sequences of effectively
instantaneous control pulses.  While such idealized assumptions must
(and can~\cite{VK}) be significantly weakened for realistic
applications, the bang-bang setting provides the most convenient
starting point for discussing stochastic schemes.

Pictorially, it is helpful to visualize a control protocol in terms of
the path that $U_c(t)$ follows in the space of unitary transformations
on $S$.  For bang-bang controls, such a path is described as a
piecewise constant time dependence, with jumps between consecutive
values corresponding to the application of an instantaneous control
kick.  In particular, a large class of decoupling schemes may be
obtained by constraining such values to belong to a discrete subgroup
$\cg$ of unitary operators, the so-called {\em decoupling
group}~\cite{VKL99}. Let $\cg =\{g_\ell\}$, where $g_\ell$, $\ell=0,
\ldots, {|\cg|-1}$, $g_0={\bf I}_S$, denote group elements~\footnote{I
am identifying an abstractly defined decoupling group with its
image under a {\em projective} representation in $\ch_S$.  Loosely
speaking, $\cg$ is a ``group up to phase factors'', in general.  This
is irrelevant for the present discussion.}.  {\em Cyclic decoupling
according to $\cg$ over $T_c$} is implemented by sequentially steering
$U_c(t)$ through each of the $|\cg|$ group elements that is,
\begin{equation}
U_c[(j -1)\Delta t +s]=g_j\:,  \hspace{3mm} s\in [0,\Delta t)\:,
\label{bb}
\end{equation}
with $\Delta t=T_c/|\cg|$ and $j=1,\ldots,|\cg|$.  One can prove 
that, in a {\em fast control limit} where
\begin{equation}
T_c \rightarrow 0\:,\:M\rightarrow\infty\:,  \hspace{3mm}
T=MT_c >0\:,
\label{fast}
\end{equation}
the leading contribution to the average Hamiltonian resulting from 
$\tilde{H}_{SE}(t)$ in Eq.~(\ref{lham}) is given by
\begin{eqnarray}
\overline{H}^{(0)}_{SE}&\hspace{-1mm}=\hspace{-1mm}& 
\overline{H}_S\otimes {\bf I}_E + {\bf I}_S\otimes H_E + 
\sum_a \overline{J}_a \otimes B_a\:, \nonumber \\
\overline{X}&\hspace{-1mm}=\hspace{-1mm}&
{1 \over T_c}\int_0^{T_c} dt \,U_c^\dagger (t) X U_c(t)\:.
\label{timeav}
\end{eqnarray}
The advantage of group-based decoupling scheme is that the above time
averages are directly mapped, via Eq.~(\ref{bb}), to averages over the
control group $\cg$, effectively implying a {\em symmetrization of the
controlled dynamics according to $\cg$}~\cite{VKL99,Zanardi99,Wocjan}.
If, in particular, the action of $\cg$ is {\em irreducible}, then by
Schur's lemma
\begin{equation}
\overline{X}={1\over |\cg|} \sum_{g_\ell \in \cg} g_\ell^\dagger 
X g_\ell ={\tr (X) \over d_S} {\bf I}_S =0\:, 
\end{equation} 
immediately implying complete decoupling as in Eq.~(\ref{dec1}).

While cyclic schemes may be very powerful and conceptually simple,
they are only applicable (at least in the simple formulation presented
here) to time-independent control systems.  Also, because averaging
requires traversing {\em all} of $\cg$, they tend to become very
inefficient as the size of $\cg$ grows.  The basic idea that underlies
{\em random decoupling according to $\cg$} is to replace sequential
cycling with {\em random sampling over ${\cg}$}. In the simplest kind of
protocols, the value of the propagator $U_c(t)$ is determined by a group
element which is picked uniformly at random in $\cg$ that is,
\begin{equation}
\mbox{Prob}\,(g_\ell) = {1\over |\cg|}\:, \hspace{3mm} 
\forall g_\ell \in \cg\:. 
\label{pick}
\end{equation}
Thus, both the past control operations and the times at which they are
effected are known, but the future control path is random.  Under
these conditions, no average Hamiltonian formulation is viable, and
averaging effects emerge through {\em ensemble} rather than {\em time}
averages,
\begin{equation}
\langle\langle X(t) \rangle\rangle ={\bf E}\left\{
U_c^\dagger (t) X(t) U_c(t)\right\} \:.
\label{randomav}
\end{equation}
Under the uniformity assumption, such expectation values again 
reduce to averages over $\cg$, leading to the possibility of
{\em stochastic averaging},
\begin{equation}
\langle\langle X(t) \rangle\rangle ={1\over |\cg|} 
\sum_{g_\ell \in \cg} g_\ell^\dagger 
X (t) g_\ell =0\:.
\end{equation}
The two key questions to address for random decoupling are to
understand whether stochastic protocols are indeed capable of
achieving decoupling and, if so, how they perform compared to
deterministic counterparts.  We focus here on the first question, by
presenting an explicit derivation of an error bound for randomized
control directly within the open-system
context~\footnote{In~\cite{VK05}, a detailed proof was obtained for
the closed-system setting, and used to sketch the main steps leading
to the open-system result.}.

\section{Random decoupling}
\subsection{General error bounds}

We begin by recalling a few preliminary facts. 

\begin{remark} Let 
$|| A ||_2=\mbox{Max}\,|\mbox{eig}\,(\sqrt{A^\dagger A})|$ denote the
operator $2$-norm of $A$.  Then (see e.g.~\cite{bhatia}) 
\\ $(i)$ \ \ $|| A ||_2 =
\mbox{Max}\,|\mbox{eig}\,(A)|\:,$ $\:\forall A=A^\dagger$; \\ 
$(ii)$ \ $||A B ||_2 \leq || A ||_2 || B ||_2\:,$ $\:\forall A, B$; \\ 
$(iii)$ If $U$ is unitary, $|| U^\dagger A U ||_2 = || A ||_2\:, $ 
$\forall A$.
\label{rk}
\end{remark}

\begin{lemma} \label{rank}
Let $A$ be any rank-$1$ operator on $\ch_S$. Then 
$$|\tr(A)|\leq || A ||_2 $$
\end{lemma}

\vspace*{2mm}

\proof $A$ may be represented as $A\simeq |v\rangle\langle v|$, for a
$d$-dimensional complex vector $|v\rangle=[v_1,\ldots, v_d]$ with norm
$||v||=\sum_k |v_k|^2$. Then 
$$ |\tr(A)|= |v_1| \leq || v || = \mbox{Max}\,
|\mbox{eig}\,(\sqrt{  |v\rangle\langle v| })| = || A||_2 \:.$$

Q.E.D. 

\begin{theorem} \label{main}

Let $S$ be an open quantum system described by a Hamiltonian of the 
form (\ref{totaldrift}). Suppose that the control protocol satisfies
the following assumptions: \\
$(i)$ \ \ ({\em Irreducibility}) $\cg$ acts irreducibly on $\ch_S$. \\
$(ii)$ \ ({\em Uniformity}) $U_c(t)$ is uniformly random for each $t$. \\
$(iii)$ ({\em Independence}) For any $t,s >0$, $U_c(t)$ and $U_c(t+s)$ 
are independent for $s > \Delta t$. 

If, in addition, the total interaction Hamiltonian is uniformly bounded 
in time,
\begin{equation}
\Big|\Big| H_S(t) \otimes {\bf I}_E + 
\sum_a J_a(t) \otimes B_a (t) \Big|\Big|_2 < k \:,\hspace{3mm}\forall t\:, 
\label{b}
\end{equation}  
then 
\begin{equation}
\epsilon_T = O \left({T \Delta t \, k^2}\right) 
\;\;\;{\mbox{for}}\;\;\; {T \Delta t \,k^2} \ll 1 \:.
\label{rbound}
\end{equation}
\end{theorem}

\vspace*{2mm}

\proof Let $\pi_S$ be an arbitrary pure state of $S$. The first step
is to cast the pure-state error probability (\ref{psep}) in a more
convenient form to bound. By purifying the initial state of $E$ if
necessary, we may assume that $\rho_{SE}(0)= \pi_S \otimes \pi_E$,
both $\pi_{S,E}$ being one-dimensional projectors.  By using the
definition of partial trace and the cyclicity property of the full
trace, we have
\begin{eqnarray}
\epsilon_T(\pi_S)&\hspace{-1mm}=\hspace{-1mm}&
{\bf E}\left\{ \tr_{S} \left( \pi^\perp_S 
\tilde{\rho}_S(T)\right)\right\} \\ 
&\hspace{-1mm}=\hspace{-1mm}& {\bf E}\left\{ \tr_{SE} 
\left( \pi^\perp_S\otimes {\bf I}_E \tilde{\rho}_{SE} (T)
\right)\right\} \nonumber \\
&\hspace{-1mm}=\hspace{-1mm}& 
{\bf E}\left\{ \tr_{SE} 
\left( \pi^\perp_S\otimes {\bf I}_E \tilde{U}(T) 
\pi_S \otimes \pi_E \tilde{U}^\dagger (T) \right)\right\}
\nonumber \\
&\hspace{-1mm}=\hspace{-1mm}&  {\bf E}\left\{ \tr_{SE} 
\left( \pi^\perp_S\otimes {\bf I}_E \tilde{U}'(T) 
\pi_S \otimes \pi_E {{\tilde{U}'}} (T)^\dagger 
\right)\right\} \nonumber
\:,
\end{eqnarray}
where the relation (\ref{relation}) has been used, and $U_E(t)$
drops. Let $H'_{SE}(t)$ denote the interaction Hamiltonian of
Eq.~(\ref{b}).  Then the task is to bound the error in implementing
identity design on the logical interaction propagator at time $T$,
\begin{equation}
\tilde{U}'(T)={\cal T} \hspace*{-.5mm}
\exp\left \{ -i \hspace{-0.5mm} \int_0^T du 
\tilde{H}'_{SE}(u) \right\} \:,
\end{equation}
with $\tilde{H}'_{SE}(t)= U_c^\dagger(t) H'_{SE}(t) U_c(t)$ given 
in Eq.~(\ref{hp}).

The above propagator may be expressed as follows:
\begin{equation}
\tilde{U}'(T)= \sum_{n=0}^\infty I_n(T)\:,
\end{equation}
\begin{equation}
I_n(T) =(-i)^n \int_{0\leq u_1\ldots\leq u_n\leq T} 
\hspace*{-3mm} d {\bf u} \,
   \tilde{H}'_{SE}({u_n}) \ldots \tilde{H}'_{SE}({u_1})\:,
\end{equation}
and similarly for $\tilde{U}(T)^\dagger$, with 
$d{\bf u}= du_1\ldots du_n$. Thus, we need to calculate 
\begin{eqnarray}
\epsilon_T&\hspace{-7mm}(\pi_S)\hspace{-6mm}&= \nonumber \\
\hspace{-6mm}&=& \hspace{-3mm} {\bf E}\left 
\{ {\tr}_{SE} \hspace*{-1mm}\left( 
   \sum_{n,m=0}^\infty \hspace*{-2mm}\pi_S \otimes \pi_E 
I_m(T)^\dagger
\pi_S^\perp\otimes{\bf I}_E I_n (T) \hspace{-1mm}\right)
\hspace{-1mm}\right\} \,.
\nonumber 
\end{eqnarray}
The contributions with $n=0$ or $m=0$ vanish because of $\pi_S^\perp$ 
and $\pi_S$ cancel each other upon exploiting the cyclicity of the 
trace. Because $\epsilon_T(\pi_S)\geq0$, 
\begin{eqnarray}
|\epsilon_T&\hspace{-7mm}(\pi_S)|\hspace{-6mm}&\leq \nonumber \\
\hspace{-6mm}&\leq& \hspace{-3mm} 
\sum_{n,m \geq 1} \hspace*{-3mm}\
\Big| {\bf E}\Big \{ {\tr}_{SE} \Big( 
\pi_S \otimes \pi_E I_m(T)^\dagger
\pi_S^\perp\otimes{\bf I}_E I_n (T)
\Big) \Big \}  \Big|\,. \nonumber
\end{eqnarray}
Under the assumption of sufficiently smooth behavior, the expectation 
may be moved under the integral. Fix a pair of integers $n,m \geq 1$,
then the relevant contribution is 
\begin{eqnarray}
\int_{{W}^{(n,m)}} \hspace{-6mm}d{\bf u}\, d {\bf t}\:
{\bf E}\left \{ \pi_S\otimes \pi_E \tilde{H}'_{SE}(t_1) \ldots 
\tilde{H}'_{SE}(t_m)
\pi_S^\perp \otimes {\bf I}_E \nonumber \right. \\
\left. \tilde{H}'_{SE}(u_n)\ldots 
\tilde{H}'_{SE}(u_1)\right\}\:,
\label{nm}
\end{eqnarray}
where the integration region $W^{(n,m)}=\{ ({\bf u}, {\bf t})\,|\,
0\leq u_1$ $\ldots\leq u_n \leq T;\: 0\leq \ldots \leq t_m\leq T\}$.
Let $W_1^{(n,m)}(\Delta t)\subset W^{(n,m)}$ denote the subset of
points satisfying that $u_\ell$, $t_\ell$ are each time-ordered and
{\em no} $u_\ell$ or $t_\ell$ is further away than $\Delta t$ from the
rest, and let $W_2^{(n,m)}(\Delta t)\subset W^{(n,m)}$ denote the
remaining region. Because, within $W_2^{(n,m)}(\Delta t)$, at least
one of the integrating variables is more than $\Delta t$ away from 
all the other variables, the independence assumption $(iii)$ allows 
the expectation relative to such a variable to be taken separately.  
By the uniformity assumption $(ii)$ on $U_c(t)$ for all $t$, and by 
the tracelessness assumption on $H'_{SE}(t)$ for all $t$, such an
expectation vanishes. Therefore, $W_1^{(n,m)}(\Delta t)$ is the 
only subset of points contributing to the expectation in
Eq. (\ref{nm}). Let $d{\bf w}^{(n,m)}$ denote the corresponding
integration measure. Then
\begin{eqnarray}
\epsilon_T(\pi_S) \leq \sum_{n,m \geq 1} \int_{W_1^{(n,m)}} 
d{\bf w}^{(n,m)}\,\hspace{3cm} \nonumber \\ 
\left| {\bf E}\left \{ {\tr}_{SE} \hspace*{-1mm}\left( 
\pi_S \otimes \pi_E \tilde{H}'_{SE}(t_1) \ldots \pi_S^\perp\otimes 
{\bf I}_E \ldots \tilde{H}'_{SE}(u_1) \right) \right\}  
\right| \nonumber \nonumber \\
\hspace*{-3cm}\leq  \sum_{n,m \geq 1} \int_{W_1^{(n,m)}}
d{\bf w}^{(n,m)}\,\hspace*{3cm}   \nonumber \\
{\bf E} \left\{ \left|  
{\tr}_{SE} \hspace*{-1mm}\left( 
\pi_S \otimes \pi_E \tilde{H}'_{SE}(t_1) \ldots \pi_S^\perp\otimes 
{\bf I}_E \ldots \tilde{H}'_{SE}(u_1)
\right) \right|\right\}  \nonumber \,,
\end{eqnarray}
where in the second step Jensen's inequality has been used. By noticing
that the argument of the trace is a rank-1 operator, Lemma~\ref{rank} 
may be used to simplify
\begin{eqnarray} 
\epsilon_T(\pi_S) \leq \sum_{n,m \geq 1} \int_{W_1^{(n,m)}} 
d{\bf w}^{(n,m)}\,\hspace{3cm} \nonumber \\ 
\hspace*{5mm}{\bf E} \left\{ \Big|\Big|  
\pi_S \otimes \pi_E \tilde{H}'_{SE}(t_1) \ldots \pi_S^\perp\otimes 
{\bf I}_E \ldots \tilde{H}'_{SE}(u_1)\Big|\Big|_2 \right\}  
\nonumber \,, \hspace{5mm}\nonumber  \\
\leq \sum_{n,m \geq 1} {\mbox{Vol}}({W_1^{(n,m)}}) k^{n+m}\:, 
\hspace{2.6cm}
\nonumber
\end{eqnarray}
where the inequality $(ii)$ in the Remark~\ref{rk} and the uniform 
bound $k$ for $H'_{SE}(t)$ in (\ref{b}) have been used, and 
{Vol}$({W_1^{(n,m)}})$ is the volume of ${W_1^{(n,m)}}$. Note that 
the dependence upon $\pi_S$ has disappeared at this point.

The above volume may be estimated through a combinatorial argument.
First, notice that given the two ordered lists $0\leq u_1
\leq\ldots\leq u_n\leq T$, $0\leq t_1 \leq\ldots\leq t_m\leq T$, there
are ${n+m\choose m}$ different merged orderings. Fix a particular
one. Then each element needs to be either within $\Delta t$ of the
next one or of the previous one. Make a choice for the odd-numbered
elements, the first element being labeled $1$.  There are at most
$2^{\lceil (n+m)/2\rceil}$ such choices.  For each of them the
contribution to the volume may be bounded by ordering the
even-numbered elements, then by inserting the odd ones, ignoring 
the ordering constraint now.  Finally,
\begin{eqnarray}
{\mbox{Vol}}({W_1^{(n,m)}}) &\hspace{-2mm}\leq \hspace{-2mm}&
{n+m\choose m} \frac{\,T^{\lfloor(n+m)/2\rfloor}
(2\Delta t)^{\lceil (n+m)/2 \rceil}}{(\lfloor (n+m)/2\rfloor)!} \nonumber \\
&\hspace{-2mm}\leq\hspace{-2mm} 
& 2^{\lceil (n+m)/2\rceil} T^{\lfloor (n+m)/2\rfloor} 
(2\Delta t)^{\lceil (n+m)/2\rceil}  \nonumber \\
&\hspace{-2mm}\equiv\hspace{-2mm} &V_{nm}\:,
\label{vnm}
\end{eqnarray}
where the inequalities ${n+m\choose m}\leq 2^{n+m-1}$ (for $n+m\geq 2$),
and $\lfloor (n+m)/2\rfloor ! \geq 2^{\lfloor (n+m)/2 \rfloor -1}$ have 
been exploited.

The last step is to sum over $n,m$:
\begin{equation}
\mbox{Max}_{\pi_S} \{\epsilon_T(\pi_S) \} \equiv
\epsilon_T \leq \sum_{n,m =1}^\infty
V_{nm} k^{n+m} \:. 
\label{fin}
\end{equation}
This may be done by considering separately the four partial sums 
where both $n$ and $m$ have the same (even or odd) parity, or they
have opposite (even-odd or odd-even) parity, respectively, and by
evaluating the $\lfloor\, \rfloor$, $\lceil \, \rceil$ in Eq. 
(\ref{vnm}) accordingly. Lengthy but straightforward calculations 
yield
\begin{equation}
\epsilon_T \leq (4T\Delta t k^2) \frac{1+ 8 
\Delta t k + 4T\Delta t k^2 }
{(1-4T\Delta t k^2)^2} = O(T \Delta t k^2) \:,
\end{equation}
for values of $T \Delta t k^2 \ll 1$, as quoted in Theorem~\ref{main}.

Q.E.D. 

\begin{remark}  By setting all the coupling operators $J_a=0$, the 
error bound for random decoupling of a closed or classically 
time-dependent control system is obtained.  
\end{remark}

According to the above Theorem, the performance of stochastic control
can be made arbitrarily high by appropriate design, in particular by
choosing a sufficiently small $\Delta t$ in the present setting.
Remarkably, this implies the possibility to {\em arbitrarily suppress
on average decoherence in the logical frame}.  Note that, unlike
deterministic decoupling, stochastic schemes place {\em no}
restriction on the time dependence of $H_0(t)$, only on the maximum
eigenvalue of the interaction part, $H'_{SE}(t)$.  The latter,
however, may diverge in physical situations involving
infinite-dimensional environments.  Thus, appropriate care is needed
to properly define the relevant strength $k$ in such
situations~\cite{KLV,terhal}.  Physically, the parameter $k^{-1}$ is
of the order of the {\em shortest} correlation time present in the
interaction to be removed. While this provides the relevant time scale
to the purposes of obtaining an {\em upper} error bound, {\em lower}
or {\em typical} error bounds may be better in specific situations,
depending on the details of both the system and the environment.

\subsection{Example: Control of a single noisy qubit}

A simple illustrative example is provided by a single two-state system
(a qubit) dissipatively coupled to a quantum reservoir.  In this case
$\ch_S={\bf C}^2$ and a basis for the traceless operators on $S$ is
given by the Pauli operators, $\sigma_\alpha$, $\alpha=x,y,z$.
Consider for simplicity a time-independent open-system dynamics.
Eq.~(\ref{totaldrift}) takes then the form
\begin{equation}
H_0 =\omega_0 \sigma_z \otimes {\bf I}_E
+ {\bf I}_S \otimes H_E + \sum_\alpha \sigma_\alpha \otimes 
B_\alpha\:, 
\end{equation}
where $\sigma_z$ represents the energy eigenbasis of the isolated
qubit, and $\omega_\alpha$, $B_a$ are appropriate real parameters and
Hermitian environment operators, respectively.  Complete decoupling
may be achieved in the deterministic setting by cycling the control
propagator through a (projectively represented) error group for the
qubit\footnote{The abstract decoupling group is
${\cal Z}_2 \times {\cal Z}_2$ in this case.} that is, 
$\cg_P\simeq \{ {\bf I}_S, \sigma_x,\sigma_y,\sigma_z\}$.  
Thus, $T_c= 4\Delta t$, and Eq.~(\ref{bb}) yields
\[ U_c(t)= \left\{ \begin{array}{ll}
{\bf I}_S & t\in \Delta t_1\:, \\
    \sigma_x & t\in \Delta t_2\:,\\ 
    \sigma_y & t\in \Delta t_3\:,\\ 
    \sigma_z & t\in \Delta t_4\:.\\ 
\end{array}\right. \]
In practice, this corresponds to a series of four equally spaced
bang-bang so-called $\pi$- (or $180^\circ$-) pulses, alternating
between the $\hat{x}$ and $\hat{z}$ axes. In terms of the control
inputs introduced in (\ref{bil}), a $\pi$-pulse along the $\alpha$
axis may be performed by applying a linearly polarized oscillating
field
\begin{eqnarray*}
H_\alpha u_\alpha(t)= 
\sigma_\alpha\,V(t) \cos[\omega (t-t_P)]\:,\hspace{9mm}\\
V(t)=V[\theta (t-t_P)-\theta (t-t_P-\tau)]\:,  \hspace{3mm} V>0\:,
\end{eqnarray*}
where $\omega=\omega_0$ on resonance, $t_P$, $\tau$ are the time 
at which the pulse is applied and its duration, respectively, and 
$2V\tau=\pi$ with $\tau\rightarrow 0$, $V\rightarrow 
\infty$ to satisfy the bang-bang requirement. 

For random decoupling over the Pauli group $\cg_P$, the control
prescription (\ref{pick}) corresponds to applying a sequence of
$\pi$-pulses with are randomly drawn from $\cg_P$ that is, each of the
Pauli operators is applied with probability 0.25 at times $t_j=j\Delta
t$, $j\in {\bf N}$.  Physically, the relevant strength parameter $k$
may be associated to the high-frequency cut-off $\omega_c$ that is
contained in the reservoir power spectrum and determines its frequency
response.  In general, however, additional time scales related to both
$\omega_0$ and the temperature affect the overall control
performance. Thus, according to the worst-case bound of
Eq.~(\ref{rbound}), decoherence suppression at time $T$ is achieved
provided $\Delta t$ is made sufficiently small with respect to
$\omega_c^{-1}$.  Remarkably, an exact solution for the stochastically
controlled dynamics may be obtained in the special case where
$B_x=B_y=0$, corresponding to pure decoherence.  A detailed analysis
of this limiting situation is reported in~\cite{SV05}.

\section{Conclusion}

I have discussed a control-theoretic formulation which explicitly
invokes random control design, and which is applicable to arbitrary
finite-dimensional, time-dependent open quantum control systems.  I
focused on random decoupling design for decoherence suppression as a
relevant case study, and showed how arbitrarily low error rates may be
achieved in principle.  Further study is needed to both explore
concrete applications of randomized schemes and assess their full
potential, as well as to integrate random design within existing
control settings.  Beside pointing to a still largely unexplored
territory in the theory and practice of quantum control, the ideas
presented here might allow to take advantage of novel perspectives, as
offered for instance by noisy quantum games~\cite{meyer} or randomized
algorithms for classical uncertain systems~\cite{Tempo}.  It is my
hope that the results presented here will prompt the control theory
community to further investigate the interplay between randomness and
coherence in quantum dynamical systems.

\vspace{1mm}
\section{Acknowledgments}

The original formulation of the random decoupling problem on which I
build here is joint work with Manny Knill.  I wish to thank both him
and Seth Lloyd for the pleasure of a longstanding collaboration, as
well as Lea Santos for her invaluable help on investigating
stochastically controlled systems and for a critical reading of the
manuscript.

\end{document}